\documentclass[reprint,superscriptaddress,aps,pra,longbibliography,twocolumn]{revtex4-2} 

\usepackage{graphicx}
\usepackage{float}
\usepackage{dcolumn}
\usepackage{amsmath}					
\usepackage{amsfonts}	 
\usepackage{amsthm}					
\usepackage{mathrsfs} 					
\usepackage{bm}				
\usepackage{bbm}
\usepackage{enumerate} 
\usepackage{physics}
\usepackage{appendix}
\usepackage{graphicx}
\usepackage{color}
\usepackage[dvipsnames]{xcolor}
\usepackage{orcidlink}

\usepackage{hyperref}
\hypersetup{
	colorlinks=true,
	linkcolor=blue,
	citecolor=blue,
	filecolor=magenta,
	urlcolor=blue
}

\usepackage{array}
\usepackage{tabularx}
\usepackage{booktabs}

\usepackage{siunitx}    % Align numbers by decimal point
\usepackage{multirow}

\newcommand{\vc}[1]{\bm{\mathrm{#1}}}

\newcommand{\mt}[1]{\mathrm{#1}}
\newcommand{\mllp}{\mt{LLP}}

\newcommand{\B}{\mt{B}}

\newcommand{\X}{\mathrm{X}}
\begin{document}

\preprint{APS/123-QED}

\title{Anyon-trions in atomically thin semiconductor heterostructures}

%%%%%%%%%%%%%%%%%%%%%%%%% Affiliations %%%%%%%%%%%%%%%%%%%%%%%%

\author{Nader Mostaan\,\orcidlink{0000-0002-9573-7608}}
\email{nader.mostaan@physik.uni-muenchen.de}
\affiliation{Department of Physics and Arnold Sommerfeld Center for Theoretical Physics (ASC), Ludwig-Maximilians-Universität München, Theresienstr. 37, D-80333 München, Germany}
 \affiliation{Munich Center for Quantum Science and Technology (MCQST), Schellingstr. 4, D-80799 München, Germany}

\author{Nathan Goldman\,\orcidlink{0000-0002-0757-7289}}
\affiliation{Laboratoire Kastler Brossel, Coll\`ege de France, CNRS, ENS-Universit\'e PSL, Sorbonne Universit\'e, 11 Place Marcelin Berthelot, 75005 Paris, France}
\affiliation{International Solvay Institutes, 1050 Brussels, Belgium}
\affiliation{Center for Nonlinear Phenomena and Complex Systems, Universit\'e Libre de Bruxelles, CP 231, Campus Plaine, B-1050 Brussels, Belgium}

\author{Ataç İmamoğlu\,\orcidlink{0000-0002-0641-1631}}
\affiliation{Institute for Quantum Electronics, ETH Z\"urich, CH-8093 Z\"urich, Switzerland}
 
\author{Fabian Grusdt\,\orcidlink{0000-0003-3531-8089}}
\affiliation{Department of Physics and Arnold Sommerfeld Center for Theoretical Physics (ASC), Ludwig-Maximilians-Universität München, Theresienstr. 37, D-80333 München, Germany}
 \affiliation{Munich Center for Quantum Science and Technology (MCQST), Schellingstr. 4, D-80799 München, Germany}

\date{\today}

%%%%%%%%%%%%%%%%%%%%%%%%% End Affiliations %%%%%%%%%%%%%%%%%%%%

%%%%%%%%%%%%%%%%%%%%%%%%% Abstract %%%%%%%%%%%%%%%%%%%%%%%%%%%%

\begin{abstract}

The study of anyons in topologically ordered quantum systems has mainly relied on edge-state interferometry. However, realizing controlled braiding of anyons necessitates the ability to detect and manipulate individual anyons within the bulk. Here, we propose and theoretically investigate a first step toward this goal by demonstrating that a long-lived, optically generated interlayer exciton can bind to a quasihole in a fractional quantum Hall state, forming a composite excitation we term an \emph{anyon-trion}. Using exact diagonalization, we show that mobile anyon-trions possess a binding energy of approximately $0.5 \, meV$, whereas static anyon-trions exhibit a binding energy of about $\! 0.9 \, meV$, that is linearly proportional to the quasi-hole's fractional charge. An experimental realization based on photoluminescence from localized interlayer excitons in a quantum twisting microscope set-up should allow for a direct optical observation of anyon-trions.

\end{abstract}

%%%%%%%%%%%%%%%%%%%%%%%%% End Abstract %%%%%%%%%%%%%%%%%%%%%%%%

\maketitle

\textit{Introduction.} --- The discovery of the integer and fractional quantum Hall effects in two-dimensional electron gases (2DEGs) under strong magnetic fields revolutionized condensed matter physics by revealing the crucial role of topology in understanding quantum phases of matter \cite{klitzing1980new,tsui1982two,laughlin1983anomalous,zhang1989effective,wen1990topological}. A key milestone in the fractional quantum Hall (FQH) regime was the identification of fractionally charged quasiparticles obeying anyonic statistics \cite{arovas1984fractional,leinaas1977theory}, providing the first experimental realization of Abelian anyons—particles unique to two dimensions and described by braid group representations. Their exotic properties and potential use in topological quantum computing \cite{nayak2008non} continue to drive efforts toward their control and detection both in solids and synthetic quantum simulators.

Experimental probes of anyons include shot noise measurements to detect fractional charge \cite{saminadayar1997observation,de1998direct,goldman1995resonant,dolev2008observation}, Fabry–Pérot interferometry to observe conductance phase slips \cite{nakamura2020direct,nakamura2023fabry}, collider setups \cite{bartolomei2020fractional}, and time-domain measurements of anyon braiding \cite{ruelle2025time}. 
Although these experimental approaches have provided ground-breaking results that reveal fractional charge and braiding phase of anyons, they lack spatial resolution~\cite{papic2018imaging} and bulk control that are crucial for applications such as topological quantum computing.

Recently, van der Waals heterostructures based on transition metal dichalcogenides (TMDs) \cite{manzeli20172d} have emerged as powerful platforms for exploring strongly correlated electronic states via optical spectroscopy \cite{wang2020correlated,xu2020correlated,huang2021correlated,jin2021stripe}, revealing phenomena such as generalized Wigner crystal formation \cite{shimazaki2021optical,smolenski2021signatures,regan2020mott,li2021imaging}, as well as integer and fractional Chern insulator states. A rapidly growing direction involves optically detecting and controlling exotic quantum Hall phases. This includes optical sensing of fractional quantum Hall (FQH) states \cite{popert2022optical}, manipulation using vortex light \cite{session2025optical}, attractive polaron spectroscopy in quantum anomalous Hall states \cite{gao2025probing}, and magnetic field dependent trion photoluminescence (PL) as a probe of Chern number and composite Fermi liquid states \cite{anderson2024trion}. Complementary theoretical work has proposed using quantum optics techniques to study and manipulate quantum Hall systems \cite{ghazaryan2017light,grass2018optical,grass2020optical,gullans2017high}. In TMDs, excitons behave as tightly bound particles, which can be modeled as mobile impurities and form a natural framework for realizing hybrid exciton–electron systems. Interestingly, the use of mobile impurities in fractional quantum Hall (FQH) systems is regarded as a compelling approach for exploring their exotic properties~\cite{zhang2014fractional,grusdt2016interferometric,grass2020fractional,munoz2020anyonic,vashisht2025chiral}.

\begin{figure*}[t]
    \centering
    \includegraphics[scale=0.23]{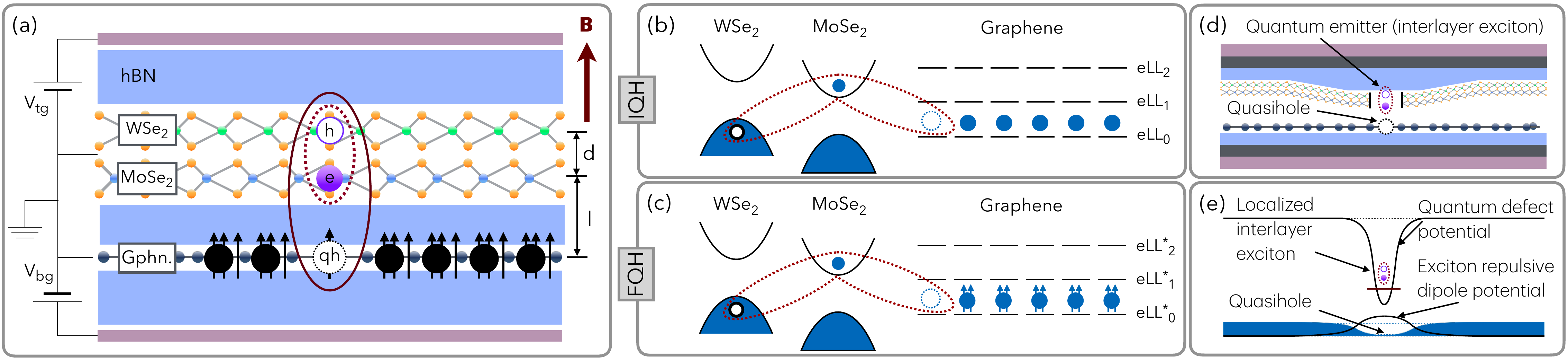}
    \caption{(a) Schematic of the device configuration to study anyon-trion formation. A $\mt{MoSe}_2/\mt{WSe}_2$ heterobilayer constitutes the optically active layer. At strong magnetic fields, the 2DEG in the proximate graphene monolayer can form quantum Hall states. An interlayer exciton in the $\mathrm{MoSe}_2/\mathrm{WSe}_2$ bilayer with repulsive interaction with the electrons in graphene can bind a hole of an IQH state (b), or a quasihole of a FQH state (c). In panels (a) and (c), a representative $\nu\!=\!1/3$ fractional quantum Hall state in the composite fermion picture is illustrated as a $\nu^{*}\!=\!1$ state of $^{2}\mt{CF}$ composite fermions. (d) An interlayer exciton as a quantum emitter on the tip of a QTM can detect quasiholes of a FQH state with nanometer-scale precision. (e) Concept of quasihole detection by a QTM: an interlayer exciton localized in a quantum defect, exerts a repulsive potential to the electrons. At the position of a quasihole—where the negative charge is reduced—the exciton experiences weaker repulsion compared to the uniform background, leading to a redshift in its energy. This redshift serves as a direct optical signature of the quasihole.}
    \label{fig:1}
\end{figure*}

In this work, we demonstrate that excitons in TMDs, interacting repulsively with electrons in a nearby quantum Hall system, can serve as sensitive probes for detecting positively-charged excitations in the bulk of incompressible integer and fractional quantum Hall states. For repulsive exciton–electron interactions, the exciton resonance exhibits a localized reduction in blueshift at the positions of (quasi)holes in (F)IQH states, indicating exciton binding to these excitations. Our proposed setup consists of a $\mathrm{MoSe}_2/\mathrm{WSe}_2$ bilayer placed near a graphene monolayer in a uniform magnetic field $B$, see Fig.~\ref{fig:1}(a). In this setup, interlayer excitons (IXs) \cite{rivera2018interlayer}, generated via non-resonant optical pumping \cite{tuugen2025optical}, act as quantum sensors for detecting localized charged excitations in graphene.

In an integer quantum Hall (IQH) state, we show that this free exciton can bind a single hole, forming a few-body composite object we refer to as a \textit{magnetic trion}, see Fig.~\ref{fig:1}(b). This few-body bound state arises uniquely from Landau level quantization; in the absence of a magnetic field, the linear dispersion of graphene prevents such binding. We find that the magnetic trion consists primarily of a $1s$ exciton scattering off a hole in graphene’s lowest Landau level.

We then consider interlayer excitons interacting with $\nu \!=\!1/3$ graphene FQH states. To capture the full many-body nature of the FQH state and its coupling to the exciton, we study small systems (up to eight electrons) on a sphere \cite{haldane1983fractional,haldane1985finite}, with the exciton confined to the same spherical surface. We employ a unitary transformation—the Lee-Low-Pines transformation~\cite{grusdt2016new,grusdt2025impurities,massignan2025polarons} to exploit the full translational and rotational symmetry of the exciton–quantum Hall system. Using this approach combined with exact diagonalization, we compute the low-energy states of the exciton-quantum Hall system and find that the exciton can bind a quasihole, forming an \textit{anyon-trion}, see Fig.~\ref{fig:1}(c). The binding energy of this state, which we estimate to be $\sim \! 0.5 \, meV$, signals the reduced charge of the quasihole. Interlayer excitons used as quantum emitters at the tip of quantum optical version of the quantum twist microscopes (QTM) allow for a sensitive and non-invasive probe of the fractional charge of the quasihole with nanometer-scale precision, see Fig~\ref{fig:1}(d,e).

\textit{Excitons in IQH states: the magnetic trion.} --- In the limit of low exciton density and hole doping ($n_{\X}, n_{h} \ll l_{B}^{-2}$ with $l_{B}\!=\!\sqrt{\hbar c/eB}$ the magnetic length), the system can be effectively described by a few-body problem involving a single exciton interacting attractively with a single hole in Landau levels. The few-body physics also plays a central role in understanding exciton-polaron formation in doped TMDs, where the exciton-polaron emerges as a coherent superposition of a bare exciton and collective trion-like excitations of the Fermi sea~\cite{Sidler2017,imamoglu2021exciton}.

\begin{figure}
    \centering
    \includegraphics[width=0.49\textwidth]{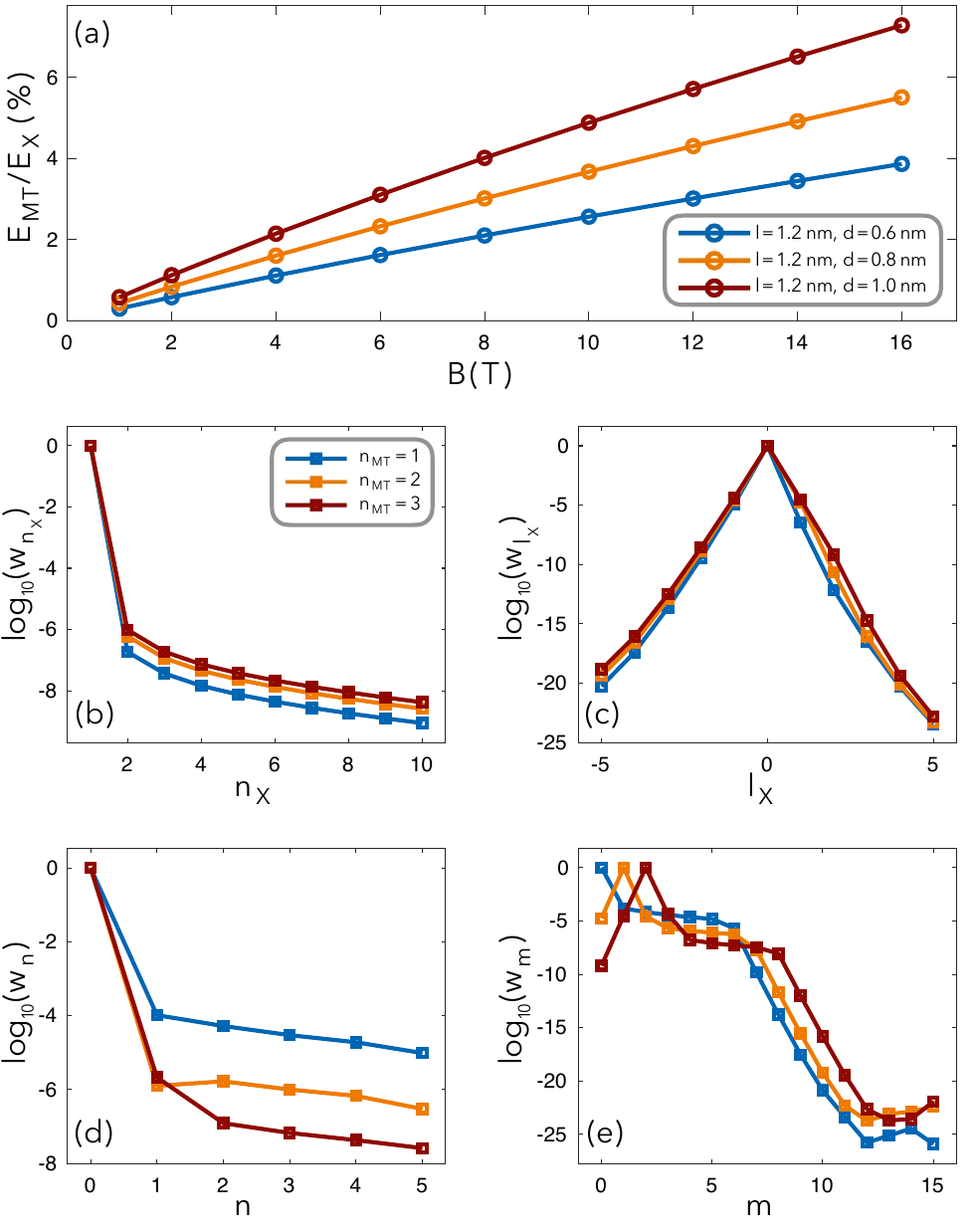}
    \caption{Few-body states of an interlayer exciton coupled to a quantum Hall hole. (a) Magnetic trion binding energy vs. magnetic field and various values of the spacing $d$ between the MoSe$_2$ and WSe$_2$ layers, for a fixed distance of $l= 1.2$~nm from tbe graphene layer. The binding energy vanishes as $B \to 0$. (b)–(e) Weights $w_{\alpha}$ for the three lowest eigenstates on a logarithmic scale, resolved by (b) exciton principal quantum number $n_{\X}$, (c) angular momentum $l_{\X}$, (d) Landau level index $n$, and (e) magnetic quantum number $m$. Weights are defined as $w_{\alpha} = \sum_{\bar{\alpha}} |\langle \alpha, \bar{\alpha} | \psi_{\mathrm{MT}} \rangle|^2$. The data show that the magnetic trion is primarily a $1s$ exciton bound to a hole in the lowest Landau level. Parameters: $B = 16\,T$, $l = 1.2,\mathrm{nm}$, $d = 0.6,\mathrm{nm}$.}
    \label{fig:2}
\end{figure}

The dynamics is governed by \textit{the magnetic trion Hamiltonian} $\hat{H}_{\mathrm{MT}} \!=\! \hat{H}_{e}+\hat{H}_{h}+\hat{H}_{Q}+\hat{V}_{eh}+\hat{V}_{eQ}+\hat{V}_{hQ}$ where $\hat{H}_{e(h)}$ and $\hat{H}_{Q}$ are the kinetic terms and $\hat{V}_{eh}$, $\hat{V}_{eQ}$ and $\hat{V}_{hQ}$ denote the corresponding interaction potentials. Hereafter, physical quantities with subscripts $e, h$, $Q$ and $\mathrm{X}$ belong to the exciton's electron, hole, the graphene hole and the exciton, respectively. To represent the few-body wave functions, it is convenient to express particle positions in terms of the trion center-of-mass $\vc{R}_{\mathrm{MT}}$, $\vc{r}_{eh} \!=\! \vc{r}_{e} - \vc{r}_h$ and $\vc{r}_{Q\mathrm{X}} \!=\! \vc{r}_{Q}-\vc{R}_{\mathrm{X}}$, where $\vc{R}_{\X}$ is the exciton center-of-mass. After a suitable gauge transformation of $\hat{H}_{\mathrm{MT}}$ \cite{suppmatLLP}, $\hat{\vc{R}}_{\mathrm{MT}}$ can be removed from the Hamiltonian. Thus, we only consider states with vanishing trion center-of-mass momentum, i.e. $\vc{P}_{\mathrm{MT}}=0$.

To obtain the bound state energy, we compare the ground state energy $E_{\rm gs}$ of $\hat{H}_{\mathrm{MT}}$ obtained via exact diagonalization, in the presence of exciton-hole interaction to the ground state $E_{\rm gs,0}$ when $V_{eQ}\!=\!V_{hQ}\!=\!0$ and define the magnetic trion binding energy by $E_{\mathrm{MT}} \!=\! E_{\rm gs,0} - E_{\rm gs}$ \cite{suppmatLLP}. The magnetic trion binding energy is depicted in Fig.~\ref{fig:2}(a) as a function of $B$, for different values of TMD bilayer spacing $d$. The binding energy increases with $B$, and asymptotically vanishes as $B \to 0$. 

To reveal the microscopic structure of the magnetic trion, we analyze the admixture of excitonic Rydberg states $(n_{\X}, l_{\X})$—principal and angular momentum quantum numbers—and graphene hole Landau level states $(n, m)$, denoting Landau level index and magnetic degeneracy, respectively. To this end, for a state $\ket{\Psi}_{\X,\mathrm{QH}}$ of the exciton-quantum Hall system and a state $\ket{\alpha}_{S}$ of the exciton ($S \!=\! \X$) or the quantum Hall ($S\!=\!\mathrm{QH}$) subsystem, we define the weight $w_{\alpha}$ by $w_{\alpha}=\sum_{\bar{\alpha}}\langle \alpha,\bar{\alpha}|\mathrm{Tr}_{\bar{S}}[\dyad{\Psi}{\Psi}]|\alpha,\bar{\alpha}\rangle$, where $\bar{\alpha}$ is the set of all the quantum numbers of $S$ other than $\alpha$, and the trace is over the subsystem $\bar{S}$ complement to $S$. The resulting weights are presented in Fig.~\ref{fig:2}(b)–(e).

Figures~\ref{fig:2}(b) and (c) reveal that the dominant excitonic contribution to the magnetic trion originates from $1s$ exciton. As shown in Fig.~\ref{fig:2}(d), the graphene hole remains predominantly in the lowest Landau level (LLL), with negligible contributions from higher Landau levels. Furthermore, for each of the three lowest trion eigenstates ($n_{\mathrm{MT}} \!=\! 1,2,3$), the wavefunction exhibits significant weight only for a single value $m$, which characterizes the orbital of the charge carrier relative to the exciton's center-of-mass. This analysis suggests a physically transparent picture: the lowest trion states can be accurately described as a bound state of a free $1s$ exciton and a Landau-quantized carrier occupying a single orbital in the LLL relative to the exciton.

This structure reflects the key interplay between the characteristic length and energy scales, and the Landau quantization. In graphene, the Landau level spacing $\hbar v_F/l_B \sim 28.2\sqrt{B[\mathrm{T}]}$ reaches $\sim 112.9 \, meV$ at $B = 16 \, T$—comparable to the interlayer exciton binding energy $E_{B,\X} \sim 100 \,meV$ \cite{van2018interlayer,rivera2018interlayer,liu2022interlayer}. Interaction terms coupling to higher excitonic states and Landau levels scale as $(a_{\X}/l_{\B})^2 \sim 0.01$ at this field, strongly suppressing such excitations. Optical excitation thus primarily creates $1s$ excitons, and the resulting few-body states remain adiabatically connected to them, with the holes in the quantum Hall state confined to the lowest Landau level. 

\textit{Excitons in FQH liquids: anyon-trions.} --- We now turn to the many-body regime in which the graphene layer hosts an electronic fractional quantum Hall (FQH) liquid at filling factor $\nu \!=\! 1/3$, see Fig.~\ref{fig:1}(c). In this setting, we demonstrate the emergence of a novel type of bound state: an anyon-trion, formed by the binding of an interlayer exciton to a quasihole of the FQH fluid. To model the coupled exciton–quantum Hall system, we employ the standard spherical geometry, which contains an exciton and $N$ electrons on the surface of a sphere of radius $R$ threaded by a magnetic flux from a monopole of strength $Q$ at its center. The full system is governed by the Hamiltonian

\begin{equation}\label{eq:HXQH}
    \hat{H} = \frac{\hat{\vc{L}}^2_{\mt{X}}}{2 M_{\mt{X}} R^2} + \sum^{N}_{i=1}\,V_{Xe}(\hat{\vc{r}}_{i}-\hat{\vc{r}}_{\mt{X}}) + \hat{H}_{\mt{FQH}} \, ,
\end{equation}
where $M_{\mt{X}}$, $\hat{\vc{r}}_{\mt{X}}$ and $\hat{\vc{L}}_{\mt{X}}$ denote the exciton's effective mass, position and angular momentum on the sphere, respectively. The term $V_{Xe}$ represents the exciton–electron interaction for the considered heterostructure \cite{wagner2025feshbach}, $\hat{\vc{r}}_i$ is the position of the $i$'th electron, and $\hat{H}_{\mt{FQH}}\!=\!\hat{\mathcal{P}}_{\mathrm{LLL}} \, \sum_{i<j}V(\hat{\vc{r}}_i-\hat{\vc{r}}_j)\,\hat{\mathcal{P}}_{\mathrm{LLL}}$ describes the interacting electron Hamiltonian projected onto the lowest Landau level via the projector $\hat{\mathcal{P}}_{\mathrm{LLL}}$.

\begin{figure}[t]
    \centering
    \includegraphics[width=0.5\textwidth]{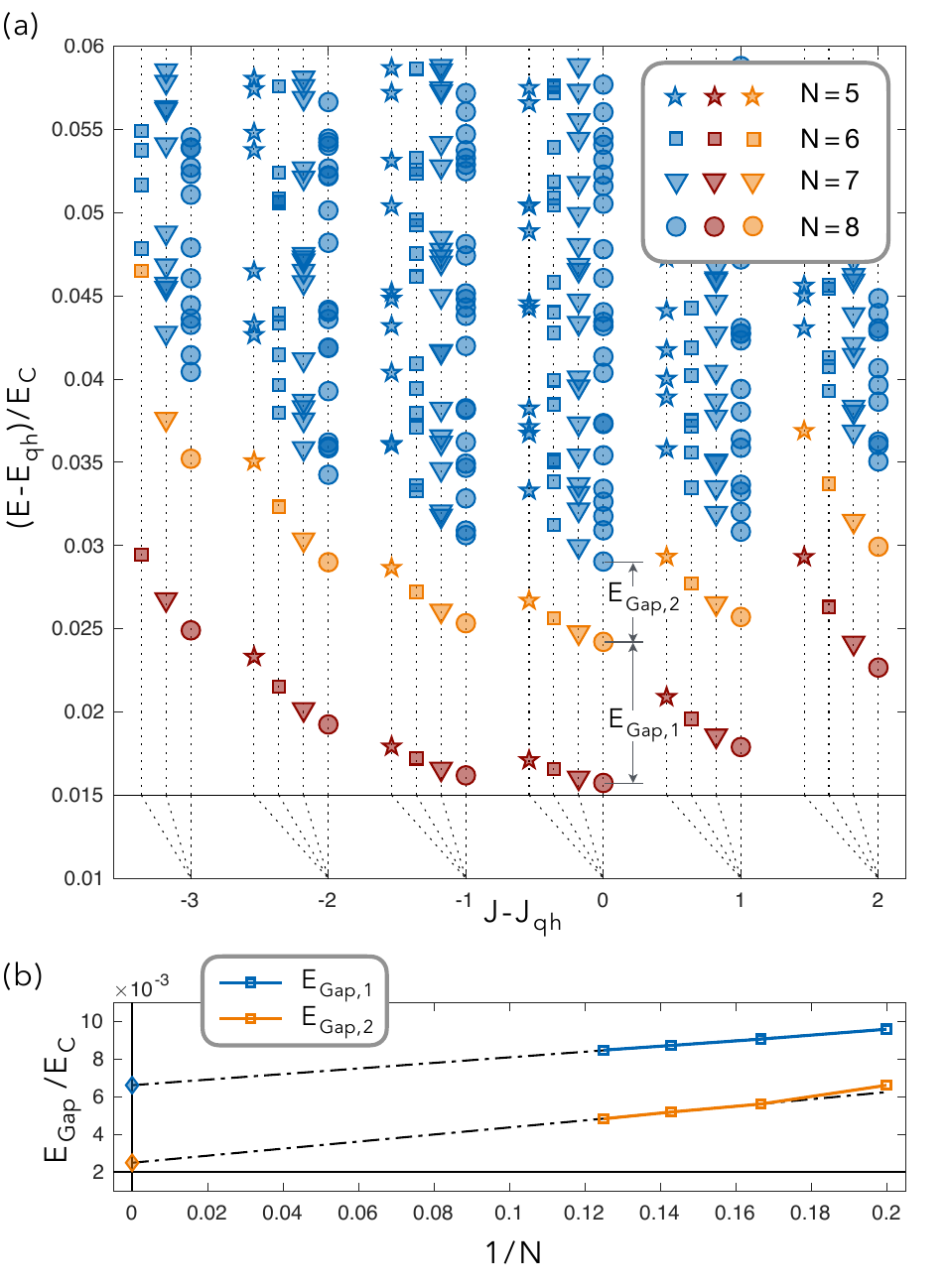}
    \caption{(a) Many-body energies, relative to the quasihole energy $E_{\mathrm{qh}}$ of the exciton interacting with a $\nu\!=\!1/3$ system of $N=5-8$ electrons at the magnetic monopole $Q_{\mathrm{qh}}$, as a function of the total angular momentum $J$ relative to the quasihole angular momentum $J_{\mathrm{qh}} \!=\! N/2$. Two energy gaps $E_{\mathrm{Gap},1}$ and $E_{\mathrm{Gap},2}$ are visible in the spectrum for different angular momenta and particle numbers. Here we take the exciton creation energy $E_{x} \!=\!0$. (b) The scaling of gaps $E_{\mathrm{Gap},1}$ and $E_{\mathrm{Gap},2}$ with $1/N$.}
    \label{fig:3}
\end{figure}

Performing exact diagonalization (ED) using a spherical geometry is a standard method for studying fractional quantum Hall systems \cite{haldane1985finite,fano1986configuration}. Introducing a mobile exciton greatly enlarges the Hilbert space, making ED computationally challenging. Instead of diagonalizing the full Hamiltonian in Eq.~(\ref{eq:HXQH}), we reduce the degrees of freedom by exploiting the system’s full rotational and translational symmetry.

This is achieved via a unitary transformation analogous to those applied in Bose polaron physics \cite{grusdt2016new,grusdt2025impurities,massignan2025polarons}, known as the Lee-Low-Pines (LLP) transformation. In spherical geometry, it takes the form \(\hat{U}_{\mllp} = \mathrm{exp}(-i\hat{\varphi}\otimes\hat{L}_{z})\,\mathrm{exp}(-i\hat{\theta}\otimes\hat{L}_{y})\,\mathrm{exp}(-i\hat{\gamma}\otimes\hat{L}_{z})\), where \((\hat{\varphi}, \hat{\theta}, \hat{\gamma})\) are exciton orientation operators and \(\hat{\vc{L}}\) is the total electronic angular momentum. \cite{suppmatLLP} Intuitively, \(\hat{U}_{\mathrm{LLP}}\) applies a rigid-body rotation into the fraction co-moving with the exciton, which is then localized at the north pole. 

After applying $\hat{U}_{\mathrm{LLP}}$, the transformed Hamiltonian $\hat{H}_{\mt{LLP}} \!=\! \hat{U}^{\dagger}_{\mathrm{LLP}} \hat{H} \hat{U}_{\mathrm{LLP}}$ is suitable for ED \cite{suppmatLLP}. Figure~\ref{fig:3}(a) shows the low-energy spectrum for $N=5-8$ electrons and a total monopole charge $Q_{\mathrm{qh}} \!=\! Q_{\mathrm{LN}}+1/2$, corresponding to a Laughlin $\nu\!=\!1/m$ FQH state at $Q_{\mathrm{LN}} \!=\! m(N-1)/2$ with a single quasihole (here $m \!=\! 3$). At $Q \!=\! Q_{\mathrm{qh}}$, the quasihole configurations form an $L \!=\! N/2$ multiplet of $N+1$ degenerate gapped ground states. Energies are measured in units of the magnetic Coulomb energy $E_{C} \!=\! e^2/4 \pi \varepsilon_0  \varepsilon_{\mathrm{hBN}} \, l_{B}$ ($\varepsilon_{\mathrm{hBN}} \!=\! 4.5$ the dielectric constant of hBN) relative to the quasihole ground state and plotted versus the total angular momentum $J$ relative to that of the isolated quasihole $J_{\mathrm{qh}} \!=\! N/2$.  Hereafter we set the parameters $B \!=\! 16 \, T$, $l \!=\! 1.2 \, nm$ and $d \!=\! 0.6 \, nm$ in our numerical calculations.

\begin{figure}[t]
    \centering
    \includegraphics[width=0.48\textwidth]{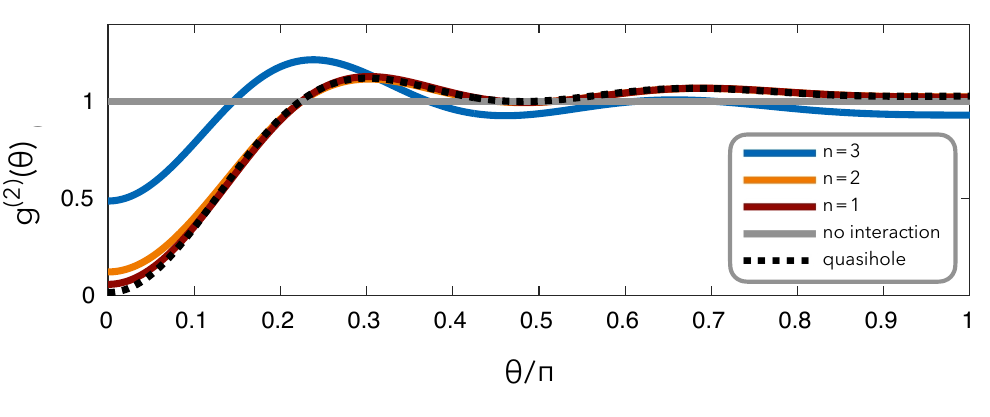}
    \caption{Exciton-electron two-particle correlation function $g^{(2)}(\theta)$ for $N\!=\!8$ electrons as a function of polar angle. The $g^{(2)}$ of a quasihole at $Q \!=\! Q_{\mathrm{qh}}$, $L_z \!=\! -N/2$ (dashed black) closely matches that of a mobile exciton repulsively interacting with electrons at the same filling and total angular momentum $J \!=\! J_{\mathrm{qh}}$ (solid red). The first excited state ($n \!=\! 2$) with an internal edge excitation, shows similar correlations (solid orange) \cite{suppmatLLP}, while the continuum onset state differs markedly (solid blue). Without exciton-electron interaction, $g^{(2)}(\theta) \!=\! 1$, indicating no spatial correlations.}
    \label{fig:4}
\end{figure}

The ground state consistently appears at $J \!=\! J_{\mathrm{qh}}$, confirming the adiabatic connection to the non-interacting ground state comprised of a zero angular momentum exciton bound to a quasihole. The quadratic center-of-mass dispersion of the anyon-trion with respect to $J-J_{\mathrm{qh}}$ evident in Fig.~\ref{fig:3}a originates from its finite mass. The spectrum exhibits two robust gaps across all $N$: $E_{\mathrm{Gap},1}$ separates the ground state from the first excited state, while $E_{\mathrm{Gap},2}$ separates the latter from the continuum. The scaling of $E_{\mathrm{Gap},1}$ and $E_{\mathrm{Gap},2}$ with $1/N$ indicates that the gaps remain finite in the thermodynamic limit, see Fig.~\ref{fig:3}(b).

To demonstrate the binding of the exciton to the quasihole and the formation of anyonic trions, a crucial quantity is the exciton-electron spatial correlation function $g^{(2)}(\vc{r})$, defined by

\begin{equation}\label{eq:g2def}
    g^{(2)}(\vc{r}) = \frac{1}{\rho N} \sum^{N}_{i=1}\, \big\langle\delta^{(2)} \big( \hat{\vc{r}}_i-\hat{\vc{r}}_{\X}-\vc{r} \big) \big \rangle \, ,
\end{equation}
where $\rho$ is the density of the uniform electron gas, and the expectation value $\langle \cdot \rangle$ is evaluated for an eigenstate  $\ket{\Psi}$ of the full system. The $g^{(2)}(\vc{r})$ function measures the probability of finding a particle in the quantum Hall system at position $\vc{r}$ relative to the exciton. 

Figure~\ref{fig:4} shows $g^{(2)}(\theta)$ for $N\!=\!8$ electrons as a function of polar angle $\theta$. Without exciton-electron interaction, $g^{(2)}(\theta) \!=\! 1$, reflecting the absence of correlations. When the repulsive interaction is included, the ground state at $Q=Q_{\mathrm{qh}}$ and $J\!=\! J_{\mathrm{qh}}$ exhibits $g^{(2)}$ closely matching that of a free quasihole at $L_{z}\!=\! -N/2$, indicating the formation of a bound state. The first excited state exhibits slightly enhanced electron density near the exciton \cite{suppmatLLP}. 

In contrast, the lowest-energy state at the onset of the continuum exhibits a pronounced enhancement of electron density at the exciton position. Analysis of its structure reveals that the weight of the $L_{z} \!=\! -N/2$ quasihole state is around $0.0037$, whereas quasihole states with $-N/2 \!<\! L_{z} \!\leq\! N/2$ account for roughly $99.16\%$ of the total weight of the quantum Hall system’s wave functions in this state. An additional $0.47\%$ originates from contributions from the gapped quantum Hall excitations. This significant admixture leads to the observed increase in local exciton density and explains the energy gap separating this state from the first excited state.

\textit{Quasihole detection using a QTM.} --- In a translationally invariant van der Waals heterostructure depicted in Fig.~1a, excitons are mobile which reduces the exciton-(quasi)hole binding due to kinetic energy cost of localizing the exciton around a (quasi)hole. A promising system to eliminate the kinetic energy cost is to use localized inter-layer excitons or quantum emitters: given that we expect (quasi)holes to be localized due to unavoidable disorder, it is important to be able to scan the localized exciton and ensure its spatial overlap with a (quasi)hole. This challenge can be addressed using a quantum optical version of the recently demonstrated quantum twisting microscope (QTM)~\cite{inbar2023quantum}. The setup we envision features a localized exciton at the microscope tip. In the limit where the exciton-electron interaction energy is small compared to the gap of the incompressible IQH and FQH state, scanning the tip over a graphene sample in high magnetic fields will show spatially uniform exciton resonances. Since quasiholes appear as local charge depletions, moving the localized exciton so that it spatially overlaps with a quasihole, will lead to a  redshift of the exciton resonance. It would then be possible to image localized quasiholes in the system (Fig.~\ref{fig:1}(c,e)).

\begin{figure}
    \centering
    \includegraphics[width=0.48\textwidth]{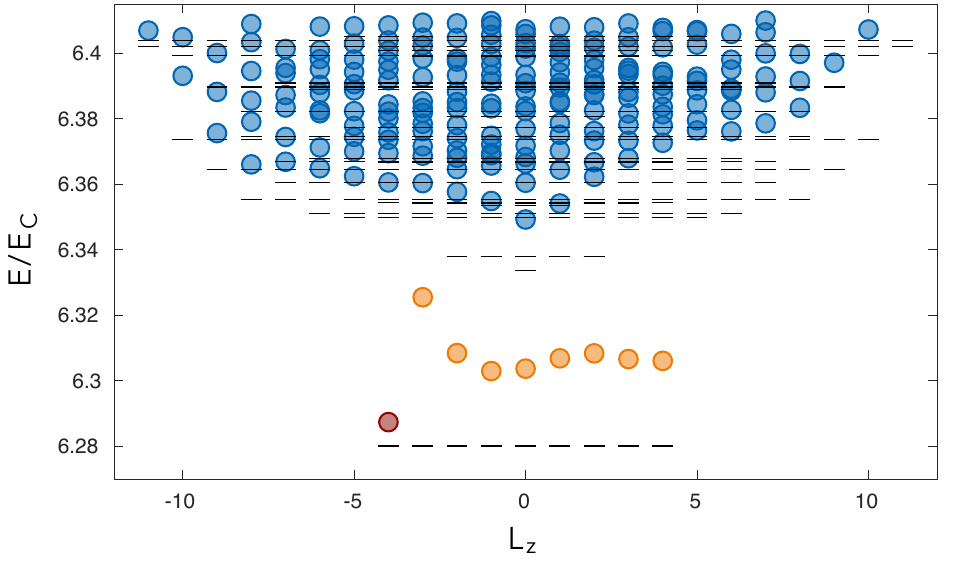}
    \caption{Low energy spectrum of a localized exciton on the tip of a QTM, in the presence of a FQH state of $N \!=\! 8$ electrons at $Q \!=\! Q_{\mathrm{qh}}$. Solid dashed lines indicate the eigenenergies of the pure quantum Hall system, where the degenerate ground state subspace corresponds to the quasihole configurations with $-N/2 \leq L_{z} \leq N/2$. The $L_z \!=\! -N/2$ state corresponds to the quasihole localized on the north pole. Circles depict the spectrum in the presence of the exciton. The bound state of the exciton and the $L_z \!=\! -N/2$ quasihole appears as the unique ground state, while other quasihole configurations form a band of states with a gap to the continuum.}
    \label{fig:8}
\end{figure}

\begin{table}[h!]
\centering
\renewcommand{\arraystretch}{1.2}
\begin{tabular}{|c|c|c|c|c|}
%\hline  <-- removed this top line
\hline
\textbf{$\boldsymbol{\nu}$} & \textbf{$\boldsymbol{N}$} & \textbf{Laughlin} & \multicolumn{2}{c|}{\textbf{Single quasihole}} \\
\cline{4-5}
 &  &  & \textbf{QTM exciton} & \textbf{mobile exciton} \\
\hline\hline
\multirow{2}{*}{$1$} 
  & 19 & 4.7550 & 1.1450 & 1.5650 \\
\cline{2-5}
  & 22 & 4.7350 & 1.1350 & 1.5550 \\
\hline\hline
\multirow{2}{*}{$1/3$} 
  & 7 & 1.2950 & 0.3700 & 0.8000 \\
\cline{2-5}
  & 8 & 1.2700 & 0.3650 & 0.7850 \\
\hline
\end{tabular}
\caption{Exciton resonance blue-shift $\Delta E(\nu,Q)$ in $meV$ defined as the blue-shift of the exciton resonance in the presence of different quantum states: IQH ($\nu \!=\! 1$) for $N \!=\! 19, \, 22$ electrons, and FQH ($\nu \!=\! 1/3$) for $N \!=\! 7, \, 8$ electrons, all at fillings corresponding to the Laughlin state $Q_{\mathrm{LN}} \!=\! (N-1)/2 \nu$ and a single quasihole state $Q_{\mathrm{qh}} \!=\! Q_{\mathrm{LN}}+1/2$. In the presence of a single quasihole, the blue-shift is reported for the localized exciton in the QTM setup as well as for the lowest energy state of the mobile exciton bound to the quasihole (anyon-trion).}
\label{tab:qh_fillings}
\end{table}

To model the QTM setup, we consider a static exciton fixed at the north pole of the sphere, corresponding to the limit $M_{\X} \to \infty$. In this setting, the $z-$axis projection of the total quantum Hall system's angular momentum, $L_{z}$, is conserved, allowing diagonalization of the full Hamiltonian within each $L_z$ sector. The resulting low energy spectrum is shown in Fig.~\ref{fig:8} for $N \!=\! 8$ and $Q \!=\! Q_{\mathrm{qh}}$. Introducing the exciton lifts the ground state degeneracy, yielding a non-degenerate gapped ground state and a band of in-gap states separated from the excitation continuum. These states, connected to the original quasihole manifold, experience an energy increase and contain an admixture of the higher excited states at the same $L_z$ of the order of $\sim 10^{-4}-10^{-3}$. Clearly, the quasihole localized on the north pole, with $L_z \!=\! -N/2$ is bound to the exciton. This highlights the QTM as a powerful, non-invasive tool capable of directly detecting quasiholes with nanometer-scale spatial resolution.

\textit{Extraction of the quasihole's fractional charge from the exciton resonance shifts} --- Exact diagonalization allows the extraction of exciton resonance blue shifts arising from repulsive interactions with electrons in a quantum Hall system. Thus, we define the blue shift, $\Delta E$, as the difference in the ground state energy of the exciton–quantum Hall Hamiltonian (Eq.~\ref{eq:HXQH}) with and without exciton–electron interaction, i.e., with $V_{Xe} \!\neq\! 0$ and $V_{Xe} \!=\! 0$, respectively. Since $\Delta E$ reflects the local electron density probed by the exciton, it depends on the filling factor $\nu$ and whether a quasihole is present. In the presence of a quasihole, the local charge depletion leads to a reduced $\Delta E$ compared to the uniform Laughlin state. In spherical geometry, introducing a single quasihole corresponds to increasing the magnetic monopole strength by $1/2$ from its Laughlin value: $Q_{\mathrm{qh}} = Q_{\mathrm{LN}} + 1/2$. We therefore denote the exciton blue shift as $\Delta E(\nu, Q)$, where $Q$ characterizes the magnetic flux and indirectly the presence of a quasihole.

In our numerical calculations, we extract $\Delta E$ at $\nu \!=\! 1$ and $\nu \!=\! 1/3$ when the quantum Hall system is either in the Laughlin ground state or hosts a single quasihole. For $N \!=\! 7,8$ particles at $\nu \!=\! 1/3$, the Laughlin state is realized by setting the magnetic monopole to $Q_{\mathrm{LN}} \!=\! 3(N - 1)/2$. To obtain the $\nu \!=\! 1$ Laughlin state, we keep the same monopole value and increase the particle number until the lowest Landau level is completely filled. Likewise, to realize a single hole at $\nu \!=\! 1$, we set the monopole to $Q_{\mathrm{qh}} \!=\! Q_{\mathrm{LN}} + 1/2$—corresponding to a quasihole on top of the $\nu \!=\! 1/3$ Laughlin state—and again increase the particle number until the system reaches $\nu \!=\! 1$ with one excess flux. Table~\ref{tab:qh_fillings} summarizes $\Delta E(\nu, Q)$ for both $\nu \!=\! 1$ and $1/3$, evaluated at Laughlin and quasihole flux values, using fixed $Q$ from the $\nu \!=\! 1/3$ configurations and the corresponding particle numbers at $\nu \!=\! 1$.

In this context, a natural definition of the anyon-trion binding energy is 
$E_{\mathrm{B},\mathrm{AT}}(\nu) \!=\! \Delta E(\nu, Q_{\mathrm{LN}}) - \Delta E(\nu, Q_{\mathrm{qh}})$. However, we find that a more accurate and experimentally robust definition is $E_{\mathrm{B},\mathrm{AT}}(\nu) \!=\! \nu \Delta E(1, Q_{\mathrm{LN}}) - \Delta E(\nu, Q_{\mathrm{qh}})$. 
This is because, at $\nu \!=\! 1/3$ and $Q_{\mathrm{LN}}$, the exciton induces weak excitations on top of the Laughlin state, which nonetheless retains approximately $92\%$ of the weight in the anyon-trion ground state. In contrast, at $\nu \!=\! 1$, such excitations are suppressed due to the absence of low-energy bulk modes, making $\Delta E(1, Q_{\mathrm{LN}})$ a more versatile reference. Linear scaling with $\nu$ takes into account the reduced electron density in the bulk at filling $\nu < 1$. Meanwhile, the exciton’s perturbation of the quasihole state is minimal: its potential is short-ranged compared to the magnetic length, and its peak coincides with the charge depletion at the quasihole center. Consequently, $\Delta E(\nu, Q_{\mathrm{qh}})$ also scales approximately linearly with $\nu$. % (see Table~\ref{tab:qh_fillings}).

Since the anyon-trion binding energy at $\nu \!=\! 1$ ($\nu \!=\! 1/3$) originates from the interaction of the charge $e$ ($e^{*}$) hole (quasihole) with the dipole field created by the exciton, we expect $E_{\mathrm{B,AT}}(\nu) \propto e(e^{*})$, respectively. Using our estimates for $E_{\mathrm{B,AT}}(\nu)$ from above we thus obtain \cite{suppmatLLP}
\begin{equation}\label{eq:fracQ}
    \frac{e^{*}}{e}= \frac{E_{\mathrm{B},\mathrm{AT}}(\nu)}{E_{\mathrm{B},\mathrm{AT}}(1)}\, .
\end{equation}
Inserting values from Table~\ref{tab:qh_fillings}, we obtain $e^{*}/e \!\simeq\! 0.3366$, within $\!\sim\! 1\%$ of the true fractional charge. The required measurement data $\Delta E(\nu,Q_{\mathrm{LN}})$, $\Delta E(\nu,Q_{\mathrm{qh}})$, and $\Delta E(1,Q_{\mathrm{qh}})$ are all accessible using QTM measurements. This demonstrates that QTM enables direct measurement of fractional charge with high accuracy and nanoscale resolution.

For mobile excitons, the observed blue shift is consistently larger due to contributions from exciton kinetic energy. This effect originates from the finite exciton mass, which induces non-trivial correlations between the exciton's motional degree of freedom and the many-body states of the quantum Hall system~\cite{suppmatLLP}. As a result, the anyon–trion binding energy is systematically smaller than the exciton–quasihole binding energy extracted in the QTM configuration. We find that this reduction in binding energy is approximately $0.42 \, meV$ at both $\nu = 1$ and $\nu = 1/3$ (see Table~\ref{tab:qh_fillings}). This kinetic energy offset leads to a significant underestimation of the fractional quasihole charge when applying Eq.~\ref{eq:fracQ}, yielding $e^{*}/e \simeq 0.2461$ for $N \!=\! 7$ and $e^{*}/e \simeq 0.2495$ for $N \!=\! 8$—a $25\%$–$26\%$ deviation from the expected value. We conclude that, although the anyon–trion constitutes a robust bound state of a mobile exciton and a quasihole, its binding energy is best interpreted as a qualitative indicator rather than a direct quantitative probe of the quasihole charge.

\textit{Conclusion and outlook.} --- We introduce an optical approach to probe quantum Hall states using interlayer excitons in TMD heterostructures. In a $\mathrm{MoSe}_2/\mathrm{WSe}_2$ bilayer above a graphene layer, excitons act as sensitive local probes for quasiholes in IQH and FQH states, ideally detectable using a  quantum optical microscope.

The emergent bound states we found in this work can serve as a foundation for developing a quantum Hall polaron theory where excitons coupled to quantum Hall fluids form quantum Hall polarons. Understanding the crossover from conventional Fermi to quantum Hall polarons as the magnetic field increases, including possible polaron-to-molecule transitions with fractionalized trions, invites new theoretical developments, such as fractionalized versions of Chevy’s ansatz \cite{chevy2006universal} or extension of the Tavis-Cummings theory of exciton-polarons \cite{imamoglu2021exciton}. A full polaron theory would also account for exciton coupling to collective modes of the FQH state, including composite fermion cyclotron and magneto-roton modes, as well as graviton-like geometric excitations \cite{haldane2011geometrical,bacciconi2025theory}.

Beyond the single-exciton regime, strong optical pumping may drive the system into collective phases—such as exciton condensates or excitonic insulators—in proximity to a quantum Hall fluid. This paves the way for realizing hybrid many-body states comprising electrons, excitons, and trions, potentially giving rise to exotic exciton–composite fermion mixtures. Such strongly interacting quantum Hall Bose–Fermi mixtures offer an exciting platform for exploring novel correlated phases of matter. Finally, we speculate that binding a quasihole to a localized exciton located at the tip of a QTM to form a localized anyon-trion could allow for braiding operation.

\textit{Acknowledgement.} --- We thank Mohammad Hafezi, Tsung-Sheng Huang, Andrey Grankin, Felix Helmrich, Haydn Adlong, Clemens Kuhlenkamp, Alperen Tüğen, Arthur Christianen and Felix Palm for insightful discussions. N. M. and F. G. acknowledge funding by the Deutsche Forschungsgemeinschaft (DFG, German Research Foundation) under Germany's Excellence Strategy -- EXC-2111 -- 390814868 and from the European Research Council (ERC) under the European Union’s Horizon 2020 research and innovation programm (Grant Agreement no 948141) — ERC Starting Grant SimUcQuam. N. G. acknowledges financial support from the ERC Grant LATIS, EOS project CHEQS, and support from the French government, administered by the National Research Agency, under the France 2030 initiative, with the reference ANR-23-PETQ-0002. A. İ. acknowledges SNSF funding under Grant Number 200021-204076.

\textit{Note added 1.} --- During completion of this work we became aware of a related work by Wagner, Neupert \cite{wagner2025sensing} which has been submitted in the same arxiv posting. 

\textit{Note added 2.} --- A preliminary version of this work was previously published in the thesis \cite{mostaan2024strongly}, including the use of the spherical Lee-Low-Pines transformation for exciton-quantum Hall systems on the sphere.

% \section*{}
\twocolumngrid
\bibliography{references}

\renewcommand{\theequation}{S\arabic{equation}}
\renewcommand{\thefigure}{S\arabic{figure}}
\renewcommand{\thetable}{S\arabic{table}}

\onecolumngrid

%\newpage

\setcounter{equation}{0}
\setcounter{figure}{0}
\setcounter{table}{0}

%%%%%%%%%%%%%%%%%%%%%%%%%%%%%%%%%%%%%%%%%%%%%%%%%%%%
\clearpage

\section*{SUPPLEMENTAL MATERIAL\\ ``Anyon-trions in atomically thin semiconductor heterostructures''}
\setcounter{page}{1}
\begin{center}
Nader Mostaan$^{1,2}$, Nathan Goldman$^{3,4,5}$, Ataç İmamoğlu$^{6}$ and Fabian Grusdt$^{1,2}$\\
\emph{\small $^1$Department of Physics and Arnold Sommerfeld Center for Theoretical Physics (ASC), Ludwig-Maximilians-Universität München, Theresienstr. 37, D-80333 München, Germany}\\
\emph{\small $^2$Munich Center for Quantum Science and Technology (MCQST), Schellingstr. 4, D-80799 München, Germany}\\
\emph{\small $^3$Laboratoire Kastler Brossel, Coll\`ege de France, CNRS, ENS-Universit\'e PSL, Sorbonne Universit\'e, 11 Place Marcelin Berthelot, 75005 Paris, France}\\
\emph{\small $^4$International Solvay Institutes, 1050 Brussels, Belgium}\\
\emph{\small $^5$Center for Nonlinear Phenomena and Complex Systems, Universit\'e Libre de Bruxelles, CP 231, Campus Plaine, B-1050 Brussels, Belgium}\\
\emph{\small $^6$Institute for Quantum Electronics, ETH Z\"urich, CH-8093 Z\"urich, Switzerland}

\end{center}

\section*{Excitons on top of an IQH state: the few-body problem }

In the following, we detail on the analysis of the exciton-IQH system few-body states. In the presence of a magnetic field, the relevant conjugate momentum of the charged particles is \textit{the kinetic momentum} $\hat{\vc{\pi}}=\hat{\vc{p}}-(q/c)\vc{A}(\hat{\vc{r}})$, where $\hat{\vc{p}} \!=\! -i\hbar \nabla_{\vc{r}}$, $q$ is the particle's electric charge ($q_{e} \!=\! -e<0$ for the electron, $q_{h} \!=\! e$ for the hole, and $q_{Q} \!=\! e$ for the graphene hole, denoted here by $Q$ to distinguish it from the exciton' hole), and $\vc{A}(\hat{\vc{r}})$ is the vector potential, which in the symmetric gauge considered in this work reads $\vc{A}(\vc{r}) \!=\! (1/2) \, \vc{B}\times \vc{r}$. The dynamics is governed by \textit{the magnetic trion Hamiltonian} $\hat{H}_{\mathrm{MT}} \!=\! \hat{H}_{e}+\hat{H}_{h}+\hat{H}_{Q}+\hat{V}_{eh}+\hat{V}_{eQ}+\hat{V}_{hQ}$ where $\hat{H}_{e(h)} \!=\! \hat{\vc{\pi}}^2_{e(h)}/2m_{e(h)}$ and $\hat{H}_{Q} \!=\! \hbar v_{F} \, \hat{\vc{\pi}}_{Q} \cdot \hat{\vc{\sigma}}$ are the kinetic terms for the electron (hole), and the graphene hole, respectively. Here we neglect the graphene valley quantum number, and only consider the sublattice pseudospin, represented by the Pauli matrices $\hat{\vc{\sigma}}$. The electron-hole, electron-$Q$, and hole-$Q$ interactions are described by $\hat{V}_{eh}$, $\hat{V}_{eQ}$ and $\hat{V}_{hQ}$, respectively. To represent the few-body states, it is convenient to choose the relative and center-of-mass coordinates of the particles, $\vc{R}_{\mathrm{MT}} \!=\! \vc{R}_{\mathrm{X}} \!=\! (m_e \vc{r}_e+m_h \vc{r}_h)/M_{\mathrm{X}}$, $\vc{r}_{eh} \!=\! \vc{r}_{e} - \vc{r}_h$, $\vc{r}_{Q\mathrm{X}} \!=\! \vc{r}_{Q}-\vc{R}_{\mathrm{X}}$, with $M_{\mathrm{X}}$ the total exciton mass. Note that the center-of-mass of the three-particle system coincides with the exciton's as a result of the vanishing mass of the graphene hole. The conjugate momenta then read $\vc{P}_{\mathrm{MT}} \!=\! \vc{p}_e+\vc{p}_h+\vc{p}_{Q}$, $\vc{p}_{eh} \!=\! (m_{h}/M_{\mathrm{X}}) \, \vc{p}_{e} - (m_{e}/M_{\mathrm{X}}) \, \vc{p}_{h}$, and $\vc{p}_{Q\mathrm{X}} \!=\! \vc{p}_{Q}$. In terms of the new coordinates and momenta, $\hat{H}_{\mathrm{MT}}$ takes a complicated form where all the canonical variables are coupled. To bring the Hamiltonian in a more tractable and physically intuitive form, we follow Ref.~\cite{gor1968contribution}, and perform a gauge transformation $\hat{U}_{g} \!=\! \mathrm{exp}\big[ \!-\! i e/\hbar c \, \vc{R}_{\mathrm{T}} \cdot \big(\vc{A}(\vc{r}_{Q\mathrm{X}}) - \, \vc{A}(\vc{r}_{eh})\big)\big]$. The gauge-transformed Hamiltonian $\hat{H}_{\mathrm{MT},g}=\hat{U}^{\dagger}_{g} \hat{H}_{\mathrm{MT}} \hat{U}_{g}$ takes the form 

\begin{equation}\label{eq:H_MTg}
\begin{split}
    \hat{H}_{\mathrm{MT},g} & = \frac{\hat{\vc{P}}^{2}_{\mathrm{MT}}}{2 M_{\mathrm{X}}} + \frac{\hat{\vc{P}}_{\mathrm{MT}}}{M_{\X}} \cdot \bigg[ \frac{2e}{c} \, \vc{A}(\hat{\vc{r}}_{eh}) - \frac{q_Q}{c} \vc{A}(\hat{\vc{r}}_{Q\X}) \bigg] 
    + \hat{H}_{\X} + \hat{H}_{Q\X} -\frac{e q_Q B^2}{2 M_{\X} c^2} \hat{\vc{r}}_{eh} \cdot \hat{\vc{r}}_{Q\X} + \frac{e^2 B^2}{8 M_{\X} c^2} \hat{\vc{r}}^2_{Q\X}
    \\ & + V_{eQ}\bigg(\frac{m_h}{M_{\X}}\hat{\vc{r}}_{eh} - \hat{\vc{r}}_{Q\X}\bigg) 
    + V_{hQ}\bigg(\frac{m_e}{M_{\X}}\hat{\vc{r}}_{eh} + \hat{\vc{r}}_{Q\X}\bigg) \, .
\end{split}
\end{equation}

In Eq.~\ref{eq:H_MTg}, the first term denotes the magnetic trion kinetic energy, the second term denotes the coupling of the magnetic trion total momentum to the internal coordinates $\vc{r}_{eh}$ and $\vc{r}_{Q\X}$, $\hat{H}_{\X} \!=\! 1/2 m_{e} \big( \hat{\vc{p}}_{eh} + e/c \, \vc{A}(\hat{\vc{r}}_{eh}) \, \big)^2 + 1/2m_h \big( \hat{\vc{p}}_{eh} - e/c \, \vc{A}(\hat{\vc{r}}_{eh}) \, \big)^2 + V_{eh}(\vc{r}_{eh})$ is the exciton's internal Hamiltonian, and $\hat{H}_{Q\X} \!=\! \hbar v_{F} \, \hat{\vc{\sigma}} \cdot \hat{\vc{\pi}}_{Q\X}$ with $\hat{\vc{\pi}}_{Q\X}=\hat{\vc{p}}_{Q\X} - q_{Q}/c \, \vc{A}(\hat{\vc{r}}_{Q\X})$. Interestingly, $\hat{\vc{R}}_{\mathrm{MT}}$ does not appear in $\hat{H}_{\mathrm{MT},g}$, enabling to restrict $\hat{H}_{\mathrm{MT},g}$ in subspaces with well-defined $\vc{\mathrm{P}}_{\mathrm{MT}}$. In this work we only consider $\vc{P}_{\mathrm{MT}}=0$. 

To obtain the few-body states and their energies, we perform exact diagonalization of $\hat{H}_{\mathrm{MT},g}$ over the basis states $\ket{n_{\X},l_{\X};n,m} \!=\! |n_{\X}, l_{\X}\rangle \otimes \, |n,m,+\rangle$ formed by the eigenstates $\ket{n_{\X},l_{\X}}$ of $\hat{H}_{\X}$, and the positive-energy eigenstates $\ket{n,m,+}$ of $\hat{H}_{Q\X}$. Here, $n_{\X}$ and $l_{\X}$ are the exciton principle quantum number and the internal angular momentum, respectively, and $|n,m\rangle$ denotes the usual Landau orbitals of a charge-$e$ particle in the symmetric gauge, with $n \geq 0$ the Landau level index and $m \geq 0$ the magnetic quantum number. To obtain the bound state energy, we compare the ground state energy $E_{\rm gs}$ of $\hat{H}_{\mathrm{MT},g}$ in the presence of exciton-carrier interaction to the ground state $E_{\rm gs,0}$ when $V_{eQ}=V_{hQ}=0$. The magnetic trion binding energy is then defined as $E_{\mathrm{MT}} \!=\! E_{\rm gs,0} - E_{\rm gs}$.

\section*{Charged particles in a uniform magnetic field in the spherical geometry}
In numerical studies of quantum Hall systems, it is a standard practice to consider spherical geometry for particles' motion \cite{jain2007composite}. In this setting, charged particles with mass $m$ and charge $q$ are constrained to move on a sphere with radius $R$, threaded by a magnetic field $\vc{B}\!=\!\big(\Phi/4 \pi R^2)\,\vc{\Omega}$, where $\Phi$ is the total magnetic flux threading the sphere and $\vc{\Omega}$ is the unit normal vector. To have single-valued wave functions, an integer number of elementary flux quanta must thread the system, thus $\Phi\!=\!(2Q)\,\phi_0$, where $\phi_0\!=\!hc/|q|$, and $Q$ is called \textit{the monopole strength} and increases in half integer steps. Similar to the planar geometry, the cyclotron motion of a charged particle in a uniform magnetic field is described by the Hamiltonian
\begin{equation}\label{eq:H_C}
    \hat{H}_{C}=\frac{\hat{\vc{\Lambda}}^2}{2 m R^2} \, ,
\end{equation}
where $\hat{\vc{\Lambda}}$ is called \textit{the kinetic angular momentum} and is related to \textit{the kinetic momentum} $\hat{\vc{\Pi}}\!=\!\hat{\vc{p}}-(q/c)\vc{A}(\hat{\vc{r}})$ by $\hat{\vc{\Lambda}}\!=\!\hat{\vc{r}}\times\hat{\vc{\Pi}}\,$.  The kinetic angular momentum is related to the spatial angular momentum $\hat{\vc{L}}$ by $\hat{\vc{L}}\!=\!\hat{\vc{\Lambda}}-\mt{sgn}(q)\,Q \,\hat{\vc{\Omega}}$. In particular, $\hat{\vc{L}}^2\!=\!\hat{\vc{\Lambda}}^2+Q^2$, since $\hat{\vc{\Omega}}\cdot\hat{\vc{\Lambda}}\!=\!0$. Thus, the eigenstates of $\hat{\vc{\Lambda}}^2$ are the same as $\hat{\vc{L}}^2$ and are given by \textit{the monopole spherical harmonics} $Y_{Q l m}(\theta,\varphi)$. The monopole spherical harmonics $Y_{Qlm}$ form an angular momentum $l$ irreducible representation of $SO(3)$ for fixed $Q$, and satisfy
\begin{equation}\label{eq:YQlmEigs}
\begin{cases}
    \, \hat{\vc{L}}^2 \, Y_{Qlm} = \hbar \, l(l+1) \, Y_{Qlm} \, , \\
    \, \hat{L}_{z} \, Y_{Qlm} = \hbar \, m\,  Y_{Qlm} \, ,
\end{cases}
\end{equation}
where $l=|Q|,\,|Q|+1,\,\cdots$ and $m=-l,l+1,\cdots,l\,$. Moreover, the $n$'th Landau level is identical to the angular momentum shell $l=|Q|+n$, with energy $E_n=(n+1/2)\,\hbar \omega_C\,$. The lowest Landau level thus has a degeneracy $2Q+1$.

To study quantum Hall systems of $N$ particles interacting with a potential $V$ with fractional lowest Landau level filling, the kinetic energy is fully quenched, and the lowest-Landau-level-projected Hamiltonian reads
\begin{equation}\label{eq:HFQH}
    \hat{H}_{\mt{FQH}} = \frac{1}{2} \sum^{\infty}_{l=0}\,V_{l}\sum^{l}_{m=-l} (-1)^m :\hat{\bar{\rho}}_{l(-m)}\,\hat{\bar{\rho}}_{lm}: \, , 
\end{equation}
where $\hat{\bar{\rho}}(\vc{\Omega})=\hat{\mathcal{P}}_{\mt{LLL}}\hat{\rho}(\vc{\Omega})\hat{\mathcal{P}}_{\mt{LLL}}$ is the LLL-projected density operator, $\hat{\bar{\rho}}_{lm}=\int\,d\Omega\,\hat{\bar{\rho}}(\vc{\Omega})\,$, and the Haldane pseudo-potentials are given by
\begin{equation}\label{eq:Vl}
    V_{l} = 2\pi \int^{\pi}_{0}\,d\theta\,\mt{sin}(\theta)\,P_{l}(\mt{cos}\,\theta)\,V\Big(2R\,\mt{sin}(\theta/2)\Big) \, .
\end{equation}

\section*{Lee-Low-Pines transformation in the spherical geometry}

Here we elaborate on the application of the Lee-Low-Pines transformation to the exciton-quantum Hall system which was crucial to obtain the results in this work. To this end, we consider the problem of an exciton as a neutral mobile impurity $X$ interacting with a quantum Hall system of $N$ electrons with mass $m$ and charge $q \!=\! -e<0$ with fractionally filled lowest Landau level. Both systems are put on a sphere with radius $R$ enclosing a magnetic monopole with strength $Q$. The full Hamiltonian reads
\begin{equation}\label{eq:Hfull}
    \hat{H} = \frac{\hat{\vc{L}}^2_{\mt{X}}}{2 M_{\mt{X}} R^2} + \sum^{N}_{i=1}\,V_{Xe}(\hat{\vc{r}}_{i}-\hat{\vc{r}}_{\mt{X}}) + \hat{H}_{\mt{FQH}} \, .
\end{equation}
In Eq.~\ref{eq:Hfull}, $M_{\mt{X}}$, $\hat{\vc{r}}_{\mt{X}}$ and $\hat{\vc{L}}_{\mt{X}}$ are, respectively, the exciton's mass, position and angular momentum, $V_{Xe}$ is the exciton-electron interaction potential, $\hat{\vc{r}}_i$ is the position of the $i$'th electron, and $\hat{H}_{\mt{FQH}}$ is given in Eq.~\ref{eq:HFQH}. The full rotational symmetry of the exciton-quantum Hall system implies that the spectrum is invariant under the rotation which bring the exciton at a position $\vc{r}_{\mt{X}}$ to the north pole $\vc{r}_{0}\!=\!R\,\vc{e}_{z}$, where $\vc{e}_{z}$ is the unit vector along the $z$ axis. This rotation is performed by the following unitary
\begin{equation}\label{eq:ULLP}
    \hat{U}_{\mllp} = e^{-i\hat{\varphi}\,\otimes\,\hat{L}_{z}}\,e^{-i\hat{\theta}\,\otimes\,\hat{L}_{y}}\,e^{-i\hat{\gamma}\,\otimes\,\hat{L}_{z} } \, .
\end{equation}
In Eq.~\ref{eq:ULLP}, $\hat{U}_{\mllp}$ acts on the whole quantum Hall system with total angular momentum $\hat{\vc{L}}\!=\!\sum^{N}_{i=1}\hat{\vc{L}}_{i}$ to rotate it by a rotation which brings a rigid body in the state $\ket{\varphi,\theta,\gamma}$ characterized with Euler angles $(\varphi,\theta,\gamma)$ to $\ket{\varphi=0,\theta=0,\gamma=0}$. Note that, although for a point-like impurity the Euler angle $\gamma$ is not defined, it should be kept throughout the calculations to maintain the entire group structure of the rotations of the quantum Hall system. 

To make use of the LLP transformation, it is necessary to understand its effect on the full Hamiltonian $\hat{H}$ as well as an arbitary state $\ket{J,M,n;\alpha}_{\mt{X,QH}}$ of the whole system where $J$, $M$ and $n$ denote angular momentum quantum numbers (we discuss the meaning of the quantum number $n$ later), and $\alpha$ denotes the rest of quantum numbers. First, we inspect different terms in the Hamiltonian. The quantum Hall Hamiltonian $\hat{H}_{\mt{FQH}}$ is obviously rotationally invariant. Application of $\hat{U}_{\mllp}$ on $V_{Xe}(\hat{\vc{r}}_{i}-\hat{\vc{r}}_{\mt{X}})$ yields
\begin{equation}\label{eq:VULLP}
    \hat{U}^{\dagger}_{\mllp}\,V_{Xe}(\hat{\vc{r}}_{i}-\hat{\vc{r}}_{\mt{X}})\,\hat{U}_{\mllp} = V_{Xe}(\hat{\vc{r}}'_{i}-\vc{r}_{0}) \, ,
\end{equation}
where $\vc{r}'_{i}=\mathcal{R}(\varphi,\theta,\gamma)[\vc{r}_{i}]$ is the rotated position of the $i$'th particle. In Eq.~\ref{eq:VULLP}, it is noticable that the action of $\hat{U}_{\mllp}$ has restricted the exciton degree of freedom to the north pole, thus instead of interacting with a mobile exciton, all the electrons in the quantum Hall system interact with a static potential localized around $\vc{r}_0$.

The action of $\hat{U}_{\mllp}$ on the exciton kinetic energy is more involved. To proceed with the analytics, it is more convenient to work with the spherical tensor components of the angular momentum operators. For any vector operator $\hat{\vc{O}}=\hat{O}_{x}\vc{e}_{x}+\hat{O}_{y}\vc{e}_{y}+\hat{O}_{z}\vc{e}_{z}\,$, the spherical tensor components are defined by  and $\hat{O}_{\mu}=(-\mu)/\sqrt{2}\,(\hat{O}_{x}+i \mu \hat{O}_{y}),\,\mu=\pm1\,$,
\begin{equation}\label{eq:OsphTens}
\begin{cases}
    \, \hat{O}_0=\hat{O}_{z} \, , & \mu = 0 \, , \\
    \, \hat{O}_{\mu}=-\frac{\mu}{\sqrt{2}}\,\big(\hat{O}_{x}+i \mu \hat{O}_{y}\big)\,, & \mu=\pm1\, .
\end{cases}
\end{equation}
Accordingly, the spherical tensor components of the exciton angular momentum operator in coordinate space read
\begin{equation}\label{eq:Lxmu}
\begin{cases}
    \, \hat{L}_{X,\,0}(\varphi,\theta,\gamma) = -i \partial_{\varphi} \, , \\
    \, \hat{L}_{X,\,\pm1}(\varphi,\theta,\gamma) = \frac{i}{\sqrt{2}}\, e^{\pm i \varphi} \bigg[ \pm \mt{cot}(\theta) \partial_{\gamma} + i \partial_{\theta} \mp \frac{1}{\mt{sin}(\theta)} \partial_{\varphi} \bigg] \, .
\end{cases}
\end{equation}
Applying $\hat{U}_{\mllp}$ to $\hat{L}_{X,\mu}$ directly gives
\begin{equation}\label{eq:LxmuLLP}
    \hat{U}^{\dagger}_{\mllp} \, \hat{L}_{X,\mu} \,  \hat{U}_{\mllp} 
    = \hat{L}_{X,\mu} - \sum_{\nu} \, D^{(1)*}_{\mu \nu}(\hat{\varphi},\hat{\theta},\hat{\gamma}) \, \hat{L}_{\nu} \, ,
\end{equation}
where $D^{(1)*}_{\mu \nu}(\varphi,\theta,\gamma)$ is the complex conjugate of the Wigner D-matrix $D^{(1)}$, and $L_{\nu}$ is the $\nu$-component of the total angular momentum of the quantum Hall system. From Eq.~\ref{eq:LxmuLLP}, it is evident that $\hat{U}_{\mllp}$ imprints the effect of the many-body medium on the exciton through the total angular momentum components $\hat{L}_{\nu}$. The LLP transformation of $\hat{\vc{L}}^2$ is then achieved by using Eq.~\ref{eq:LxmuLLP} as the following,
\begin{equation}\label{eq:L2LLP}
    \hat{U}^{\dagger}_{\mllp} \, \hat{\vc{L}}^2_{\mt{X}} \,  \hat{U}_{\mllp} = \hat{\vc{L}}'^{\,2}_{\mt{X}} - \sum_{\mu} \, (-1)^{\mu} \hat{L}'^{(-\mu)}_{\mt{X}} \, \hat{L}_{\mu} + \hat{\vc{L}}^{2} = \Big( \hat{\vc{L}}'^{(c)}_{\mt{X}} - \hat{\vc{L}} \Big)^2 \, .
\end{equation}
In Eq.~\ref{eq:L2LLP}, $\hat{\vc{L}}'^{(c)}_{\mt{X}} = \sum_{\mu} \hat{L}'^{\mu}_{\mt{X}}\,\vc{e}_{\mu}$ denotes the \textit{contravariant angular momentum} of the exciton~\cite{varshalovich1988quantum}, where $\hat{L}'^{\mu}_{\mt{X}} = (-1)^{\mu} \hat{L}'_{\mt{X}, -\mu}$. Here, $\hat{L}'_{\mt{X},\mu}$ represents the exciton angular momentum in the \textit{body-fixed frame}, defined as the frame obtained by rotating the space-fixed frame via Euler angles $(\varphi, \theta, \gamma)$. Although the term "body-fixed" traditionally refers to a frame attached to a rigid body in three dimensions, we adopt it here to refer to a frame whose $z$-axis passes through the exciton position. This usage serves as a natural generalization in the context of exciton motion on the sphere. The components of $\hat{\vc{L}}'_{\X}$ and $\hat{\vc{L}}_{\X}$ are related to one another by
\begin{equation}\label{eq:LpL}
\begin{cases}
    \, \hat{L}'^{\mu}_{\mt{X}} = \sum_{\nu} \, D^{(1)}_{\nu \mu}(\varphi,\theta,\gamma) \, \hat{L}_{\mt{X},\nu} \, , \\
    \, \hat{L}_{\nu,\mt{X}} = \sum_{\mu} \, D^{(1)*}_{\nu \mu}(\varphi,\theta,\gamma) \, \hat{L}'^{\mu}_{\mt{X}} \, .
\end{cases}
\end{equation}
In order to come from Eq.~\ref{eq:LxmuLLP} to Eq.~\ref{eq:L2LLP}, the following identities that are the defining differential equations for $D^{(j)}_{\mu \lambda}(\hat{\varphi},\hat{\theta},\hat{\gamma})$ must be used,
\begin{equation}\label{eq:CommRel}
    \begin{cases}
    [\hat{L}_{\nu},D^{(j)}_{\mu\lambda}(\hat{\varphi},\hat{\theta},\hat{\gamma})]=(-1)^{1+\nu}\sqrt{j(j+1)}\, C^{j,\mu-\nu}_{j,\mu;1,-\nu}\,D^{(j)}_{(\mu-\nu)\lambda}(\hat{\varphi},\hat{\theta},\hat{\gamma}) \, , \\
    [\hat{L}'^{\nu},D^{(j)}_{\mu\lambda}(\hat{\varphi},\hat{\theta},\hat{\gamma})]=-\sqrt{j(j+1)}\,C^{j,\lambda+\nu}_{j,\lambda;1,\nu}\,D^{(j)}_{\mu(\lambda+\nu)}(\hat{\varphi},\hat{\theta},\hat{\gamma}) \, ,
\end{cases}
\end{equation}
where $C^{j,\mu}_{j_1,\mu_1;j_2,\mu_2}=\braket{j_1,\mu_1;j_2,\mu_2}{j,\mu}$ are Clebsch-Gordan coefficients. 

The next step is to find the form an arbitrary state $\ket{J,M,n;\alpha}_{\mt{X,QH}}$ of the full system with angular momentum quantum numbers $J$, $M$ and $n$, and the rest of the quantum numbers summarized in $\beta$. To this end, we note that $\ket{J,M,n;\alpha}_{\mt{X,QH}}$ takes the most general form
\begin{equation}\label{eq:ArbSt}
    \ket{J,M,n;\alpha}_{\mt{X,QH}} = \sum_{j,L,\beta} c_{j,L;\alpha,\beta} \, \sum_{m,N}\, C^{J,M}_{j,m;L,N}\,\ket{j,m,n}_{\mt{X}}\otimes\ket{L,N;\beta}_{\mt{QH}} = \sum_{j,L,\beta} c_{j,L;\alpha,\beta} \ket{j,L;\beta}_{\mt{X,QH}} .
\end{equation}
In Eq.~\ref{eq:ArbSt}, $\ket{j,m,n}_{\mt{X}}$ is an angular-momentum-$j$ eigenstate of $\hat{\vc{L}}^2_{\mt{X}}$, and the quantum numbers $m$ and $n$ denote the angular momentum projections along the $z$-axis in the space-fixed and body-fixed frames, respectively. More explicitely,
\begin{equation}\label{eq:jmn}
    \begin{cases}
    \, \hat{\vc{L}}^2_{\mt{X}}\ket{j,m,n}_{\mt{X}} = \hbar \, j(j+1) \ket{j,m,n}_{\mt{X}} \, , \\
    \, \hat{L}_{\mt{X},z}\ket{j,m,n}_{\mt{X}} = \hbar \, m \ket{j,m,n}_{\mt{X}} \, , \\
    \, \hat{L}'_{\mt{X},z}\ket{j,m,n}_{\mt{X}} = \hbar \, n \ket{j,m,n}_{\mt{X}} \, . \\
\end{cases}
\end{equation}
The state $\ket{L,N;\beta}_{\mt{QH}}$ is an eigenstate of $\hat{H}_{\mt{FQH}}$ with angular momentum quantum numbers $L$ and $N$, and labeled by $\beta$. It is clear from the form of $\ket{J,M,n;\alpha}_{\mt{X,QH}}$ in Eq.~\ref{eq:ArbSt} that it suffices to evaluate $\hat{U}^{\dagger}_{\mllp}$ on $\ket{j,m,n}_{\mt{X}}\otimes\ket{L,N;\beta}_{\mt{QH}}$,
\begin{equation}\label{eq:ArbStLLP}
    \hat{U}^{\dagger}_{\mllp} \ket{j,m,n}_{\mt{X}}\otimes\ket{L,N;\beta}_{\mt{QH}} = \int^{2\pi}_{0}d\gamma\,\int^{\pi}_{0}d\theta\,\mt{sin}(\theta)\int^{2\pi}_{0}d\varphi \,\sqrt{\frac{2j+1}{8 \pi^2}} \, D^{(j)*}_{mn}(\varphi,\theta,\gamma) \ket{\varphi,\theta,\gamma}_{\mt{X}}\otimes\,\hat{U}^{\dagger}(\varphi,\theta,\gamma)\ket{L,N;\beta}_{\mt{QH}}\, .
\end{equation}
To proceed, we use the transformation of angular momentum representations under arbitrary rotations
\begin{equation}\label{eq:ULN}
    \hat{U}^{\dagger}(\varphi,\theta,\gamma)\ket{L,N;\beta}_{\mt{QH}}=\sum_{N'}\,D^{(L)*}_{NN'}(\varphi,\theta,\phi)\ket{L,N';\beta}_{\mt{QH}} \, ,
\end{equation}
together with several identities involving angular momentum summations and symmetry properties of the Wigner D matrices and Clebsch-Gordan coefficients to finally arrive at 
\begin{equation}\label{eq:UdagLLPjlb}
    \hat{U}^{\dagger}_{\mllp}\ket{j,L;\beta}_{\mt{X,QH}} = \sum_{N'} \, (-1)^{L+N'} \, C^{j,(-n)*}_{J,-(n+N');L,N'} \,\ket{J,M,n+N'}_{\mt{X}} \otimes \ket{L,N';\beta}_{\mt{QH}} \, .
\end{equation}
Inserting Eq.~\ref{eq:UdagLLPjlb} into Eq.~\ref{eq:ArbSt}, 
\begin{equation}\label{eq:UdagJMa}
    \hat{U}^{\dagger}_{\mllp}\ket{J,M,n;\alpha}_{\mt{X,QH}} = \sum_{L,N,\beta} \, f_{L,N,n;\alpha,\beta} \, \ket{J,M,n+N}_{\mt{X}}\otimes\ket{L,N;\beta}_{\mt{QH}} \, ,
\end{equation}
where $f_{L,N,n;\alpha,\beta}=\sum_{j} (-1)^{L+N} C^{j,(-n)*}_{J,-(n+N);L,N}\,c_{j,L;\alpha,\beta}\,$. 

Similar results have been obtained in Ref.~\cite{schmidt2016deformation} in the context of angulons. However, the underlying physics differs significantly: angulon systems involve quantum rotors with internal rotational degrees of freedom in 3D space, while in our case, angular momentum arises from the translational motion of excitons and electrons confined to 2D space. These distinctions lead to fundamentally different thermodynamic limits. Notably, in the angulon model, the quantum number $n$ is fixed at zero, whereas in the impurity–quantum Hall system on a sphere, $n$ reflects genuine angular momentum excitations of the quantum Hall fluid around the impurity.

Our exact diagonalization analysis reveals that the ground state of the anyon–trion corresponds to $n \!=\! 0$, while the first excited state is characterized by $n \!=\! 1$. This indicates that, in the first excited state, the quantum Hall system carries one additional unit of angular momentum around the exciton. This excitation directly corresponds to the first excited state identified in Ref.~\cite{munoz2020anyonic}, in the context of a heavy impurity bound to a quasiholes of atomic Laughlin states, which can be understood as an internal edge excitation of the the quasihole around the impurity.

\section*{Impact of the Exciton Mass on Exciton–Electron Correlations in Anyon–Trion Bound States}

Here, we demonstrate that the finite mass of an exciton introduces significant correlations between the exciton's motional degrees of freedom and the many-body states of the surrounding quantum Hall system. Beyond spatial exciton–electron correlations, the finite mass induces nontrivial structure in the joint wavefunction of the coupled system. Intuitively, a lighter exciton mass enhances correlations with the background medium—in this case, the fractional quantum Hall (FQH) liquid.

To quantify these correlations, we define an exciton–electron correlation energy analogous to the correlation energy in Density Functional Theory (DFT). For a given eigenstate $\ket{\Psi}$ of the full exciton–quantum Hall system, we define the correlation energy as the deviation of its exact energy $E_\Psi$ from a mean-field estimate:
\(
E_{\mathrm{MF}} = \langle \hat{H}_{\mathrm{LLP}} + \hat{H}_{\mathrm{FQH}} \rangle + V_{\mathrm{Hartree}} \, ,
\)
where the Hartree shift $V_{\mathrm{Hartree}}$ captures the exciton–electron density–density interaction energy, given by
\begin{equation}
\label{eq:VHertree}
    V_{\mathrm{Hartree}} = \sum_{l,m} V_{Xe,l} \, Y^{*}_{lm}(\hat{\mathbf{e}}_{z})\, |\langle \hat{\mathbf{e}}_z | \Psi \rangle|^2 \left\langle \sum_{i} Y_{lm}(\boldsymbol{\Omega}_i) \right\rangle \, ,
\end{equation}
where $V_{Xe,l}$ are the multipole coefficients of the exciton–electron potential, $Y_{lm}$ are spherical harmonics, and $\boldsymbol{\Omega}_i$ denotes the position of the $i$-th electron on the sphere.

\begin{figure}[t]
    \centering
    \includegraphics[width=0.6\textwidth]{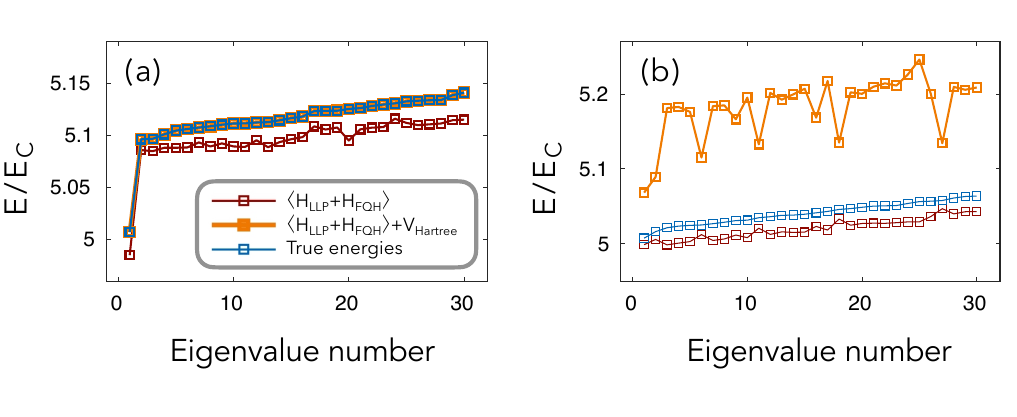}
    \caption{Comparison between exact eigenenergies of the exciton–quantum Hall system (blue), mean-field estimates $E_{\mathrm{MF}}$ (orange), and the expected values of the non-interacting Hamiltonian (red). Results are shown for the 30 lowest eigenstates of an exciton interacting whith (a) a $\nu \!=\! 1/3$ Laughlin state with $Q \!=\! Q_{\mathrm{LN}},\, J\!=\!0$, and (b) a $\nu \!=\! 1/3$ state with a single quasihole, $Q \!=\! Q_{\mathrm{qh}},\, J\!=\!J_{\mathrm{qh}}$.}
    \label{fig:5}
\end{figure}

Figure~\ref{fig:5} presents exact diagonalization results for $N \!=\! 7$ electrons with interlayer parameters $l \!=\! 1.2 \, \mathrm{nm}$ and $d \!=\! 0.6 \, \mathrm{nm}$ and $B \!=\! 16 \, T$. Panel (a) corresponds to a Laughlin state ($Q = Q_{\mathrm{LN}}$) where the system exhibits isotropy and total angular momentum $J \!=\! 0$. In this case, the exciton contributes a purely repulsive Hartree shift, well described by mean-field theory. In contrast, panel (b) illustrates the situation in the presence of a quasihole, where the many-body eigenstates of the interacting system lie significantly below the mean-field estimate, indicating the presence of strong exciton–electron correlations.

To further investigate the role of the exciton's finite mass, we examine the evolution of the first energy gap as a function of the exciton mass. We introduce a dimensionless scaling parameter $0 < s < 1$, defining a scaled mass $M_{\mathrm{X}}(s) = M_{\mathrm{X}}/s$, and correspondingly scale the kinetic contribution as $\hat{H}_{\mathrm{LLP}}(s) = s \hat{H}_{\mathrm{LLP}}$. This procedure effectively interpolates between a static and a mobile exciton, allowing us to isolate the impact of kinetic motion.

\begin{figure}[h!]
    \centering
    \includegraphics[width=0.55\textwidth]{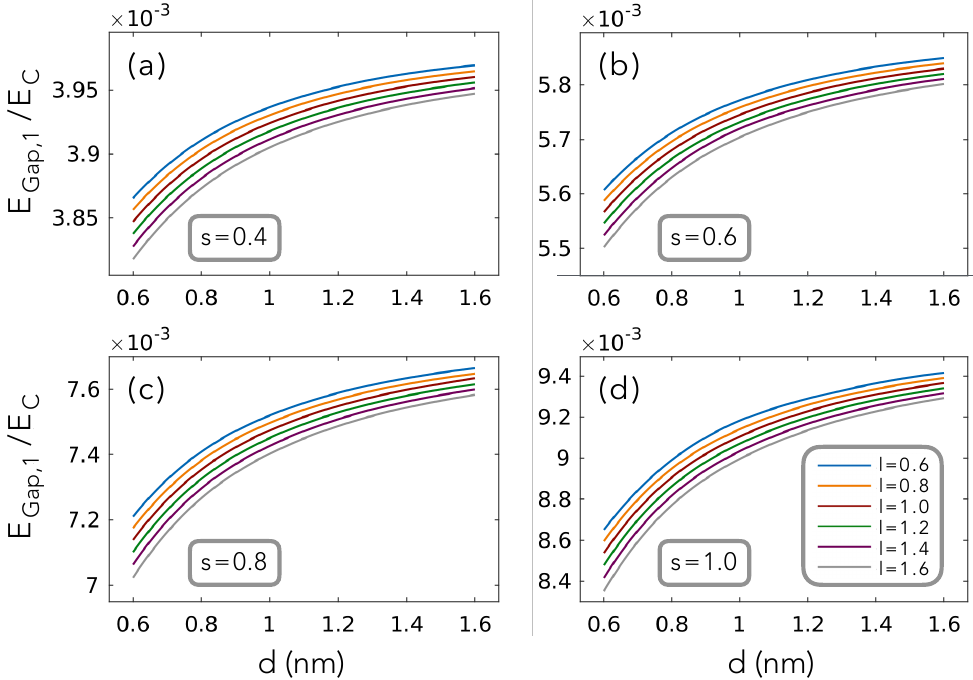}
    \caption{First excitation gap $E_{\mathrm{Gap},1}$ as a function of exciton mass scaling $s$, interlayer spacing $l$, and bilayer separation $d$. Panels (a)-(d) show that decreasing $s$ (i.e., increasing $M_{\mathrm{X}}$) reduces $E_{\mathrm{Gap},1}$, indicating weaker exciton–electron correlations. The gap is more sensitive to variations in $d$ than $l$, reflecting the stronger dependence on the exciton dipole strength. Calculations are for $N \!=\! 8$ electrons, $Q \!=\! Q_{\mathrm{qh}}$, $J \!=\! J_{\mathrm{qh}}$, and $B = 16 \, \mathrm{T}$.}
    \label{fig:6}
\end{figure}

As shown in Fig.~\ref{fig:6}, the first excitation gap $E_{\mathrm{Gap},1}$ increases with increasing contribution of the exciton kinetic term, underscoring the importance of exciton mobility in generating many-body correlations. The dependence of the gap on the bilayer separation $d$ is particularly pronounced, as it controls the dipole strength of the exciton and, hence, the magnitude of the repulsive interaction. 

\section*{Extraction of the fractional charge from anyon-trion binding energy}

To motivate the expression used in Eq.~\ref{eq:fracQ} for extracting the fractional charge from the anyon--trion binding energy, we consider the behavior of a static exciton bound to an unperturbed quasihole in a $\nu = 1/3$ fractional quantum Hall (FQH) state. Given that the exciton potential is short-ranged compared to the magnetic length, it primarily probes the electron density near the quasihole core. In this regime, the local electron density scales as $\sim \nu\, r^2$, where $r$ is the radial distance from the quasihole center. As a result, the exciton blue shift $\Delta E(\nu, Q_{\mathrm{qh}})$ appearing as the mean-field energy shift of the exciton is expected to scale linearly with $\nu$, which is also confirmed by numerical calculations. Consequently, the anyon-trion binding energy, defined as
$E_{\mathrm{B},\mathrm{AT}}(\nu) = \nu \Delta E(1, Q_{\mathrm{LN}}) - \Delta E(\nu, Q_{\mathrm{qh}})$ also exhibits a linear dependence on $\nu$. This linearity makes $E_{\mathrm{B},\mathrm{AT}}(\nu)$ a sensitive and reliable probe of the quasihole’s fractional charge.

\end{document}